\documentclass[twocolumn,prb,aps,showpacs,preprintnumbers,amsmath,amssymb,showkeys,floatfix]{revtex4-1}
\usepackage{graphicx}
\usepackage{dcolumn}
\usepackage{bm}
\usepackage{float}
\usepackage{hyperref}
\begin{document}

\title{Stability of edge states and edge magnetism in graphene nanoribbons}

\author{Jens Kunstmann,$^{1,*}$ Cem \"{O}zdo\u{g}an,$^{2}$ Alexander Quandt,$^{3,4}$ and Holger Fehske$^{3}$}
\affiliation{$^{1}$Institute for Materials Science, TU Dresden, Hallwachstr.~3, 01069 Dresden, Germany}
\email[Corresponding Author, Electronic address: ]{jens.kunstmann@tu-dresden.de}
\affiliation{$^{2}$Department of Computer Engineering, \c{C}ankaya University, Balgat, 06530 Ankara, Turkey}
\affiliation{$^{3}$Institut f{\"u}r Physik der Universit{\"a}t Greifswald, Felix--Hausdorff-Str.\ 6, 17489 Greifswald, Germany}
\affiliation{$^{4}$School of Physics and DST/NRF Centre of Excellence In Strong Materials, University of the Witwatersrand, Wits 2050, South Africa}

\date{\today}

\begin{abstract}
We critically discuss the stability of edge states and edge magnetism in zigzag edge graphene nanoribbons (ZGNRs). We point out that magnetic edge states might not exist in real systems, and show that there are at least three very natural mechanisms -- edge reconstruction, edge passivation, and edge closure -- which dramatically reduce the effect of edge states in ZGNRs or even totally eliminate them. Even if systems with magnetic edge states could be made, the intrinsic magnetism would not be stable at room temperature. Charge doping and the presence of edge defects further destabilize the intrinsic magnetism of such systems.
\end{abstract}



\maketitle

\begin{figure}[b]
\includegraphics[width=\columnwidth,angle=0]{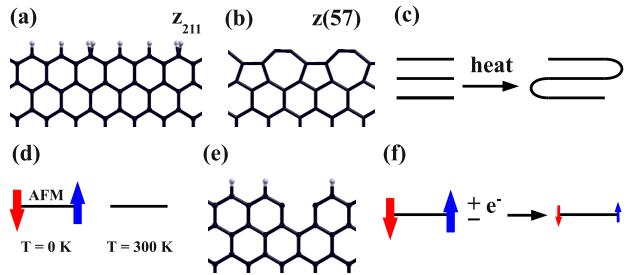}
\caption{\label{fig:outline}
Graphical outline of this article.
Top: Three mechanisms that strongly affect the edge states in ZGNRs: (a) the atomic structure of the $z_{211}$ \textit{passivated edge}, (b) the atomic structure of the $zz(57)$ \textit{reconstructed edge}, (c) \textit{edge closure}.
Bottom: The edge magnetism in (d) ideal ZGNRs is \textit{not stable at room temperature}. The low temperature magnetism is destabilized (or even totally disappears) whenever the system has (e) \textit{edge defects} or (f) is \textit{charge doped}.
}
\label{fig:overview}
\end{figure}

\section{Introduction} 
Graphene is a single layer of graphite.
Since the first successful isolation of mono- and multi-layers of graphene, \cite{novoselov_2004_sci} this material has raised tremendous attention within the scientific community. This is mainly due to the fact that graphene is an almost perfect two-dimensional system with unique electronic properties, that might ultimately lead to a new generation of nanoelectronic devices.\cite{castro_2009_rmp,quandt_2008_nt}

Graphene nanoribbons (GNRs) are one-dimensional stripes ``cut'' from graphene, which are a few nanometers in width and quasi infinite in the perpendicular direction. Initially GNRs were primarily discussed in the theoretical literature,\cite{fujita_1996_jpsj,nakada_1996_prb} but meanwhile GNRs can be fabricated and studied in the laboratory.\cite{han_2007_prl,li_2008_sci,wang_2008_prl,tapaszto_2008_natnt}
GNRs with zigzag edges (ZGNRs) are of particular interest, because they are forming \textit{edge states}.\cite{fujita_1996_jpsj,nakada_1996_prb}
The latter are localized electronic statesthat decay exponentially towards the center of the ribbon. \cite{castro_2009_rmp} The decay lengths are in the range of a few nanometers. \cite{niimi_2006_prb}
Such edge states do not only exist in perfect ZGNRs, but in any graphene system that has zigzag edge segments. \cite{stein_1987_jacs}
The localized nature of the edge states in ZGNRs gives rise to flats bands (i.e.\ parts of bands with little dispersion) in the electronic band structure, and thus to a pronounced peak in the electronic density of states (DOS) right at the Fermi level ($E_\mathrm{F}$).
This peak has been measured in the local DOS of monoatomic zigzag step edges of \textit{graphite} by scanning tunneling microscopy. \cite{kobayashi_2005_prb,kobayashi_2006_prb,niimi_2006_prb} In those measurements the peak in the  DOS did not appear right at $E_\mathrm{F}$, but very close to it. However, no such measurements for a mono-layer of graphene or GNRs are known up to date.

According to numerous theoretical studies based on density functional theory or on the mean-field Hubbard model, \cite{fujita_1996_jpsj,yazyev_2010_rpp} it is believed that the peak in the DOS at $E_\mathrm{F}$ gives rise to a magnetic instability, where the edge states become spin-polarized. In the magnetic ground state a band gap at $E_\mathrm{F}$ is opened, and the atoms are ferromagnetically ordered along one edge, and antiferromagnetically ordered between opposite edges [see Fig.~\ref{fig:spindens}(a)]. This antiferromagnetic ground state is consistent with the Lieb theorem for a bi-particle lattice within a Hubbard model type of description. \cite{lieb_1989_prl}
Potential applications of edge magnetism for spintronics are widely discussed (for an overview see [\onlinecite{yazyev_2010_rpp}]). But up to now, the edge magnetism in ZGNRs was never directly observed in local probe microscopy. Only an indirect observation has been reported quite recently. \cite{joly_2010_prb}

In this article we critically discuss the present state of research on graphene edge states and edge magnetism. We point out that magnetic edge states might not exist in real systems. And even if they do, the intrinsic magnetism would be so weak, that realistic applications in spintronic devices will not be feasible.
In order to use a physical effect in a practical device, the effect must be (i) a strong intrinsic effect and (ii) it must be stable at room temperature. We show that graphene edge magnetism does not fulfill these two criteria: (i) the effect is easily destroyed by a multitude of mechanisms and (ii) it is not stable at room temperature, even in perfect samples.

\section{Computational Methods} 
Our electronic structure calculations are based on density functional theory (DFT) \cite{kohn_1965_pr}, using a plane wave basis set (cutoff energy: 450 eV)
and the projector augmented-wave (PAW) method \cite{blochl_1994_prb_2} (we use the program \texttt{VASP},  version 4.6). \cite{kresse_1996_prb,kresse_1999_prb}
The exchange-correlation interactions were approximated by the generalized gradient approximation (GGA) within the Perdew-Wang parametrization (PW91). \cite{perdew_1992_prb_1} Total energies in the self-consistency cycle were converged such that energetic changes were less than $10^{-4}$ eV.
The $k$-space integrations were carried out using the method of Methfessel and Paxton \cite{methfessel_1989_prb} in first order, with a smearing width of 0.1 eV for the structural optimizations, and with the improved tetrahedron method \cite{blochl_1994_prb_1} for a final static calculation of the total energies.
The optimal sizes of the k-point meshes for different systems were individually converged, such that changes in the total energy were reduced to at least 3 meV/atom.
\footnote
{
Optimized k-point meshes for the different systems are: 10x10x3 (graphene), 1x1x1 (H$_2$), 5x2x3 (for all ZGNRs),  25x25x25 (bcc iron), 15x15x3 (NiO).
}

\begin{figure}[tb]
\includegraphics[width=\columnwidth,angle=0]{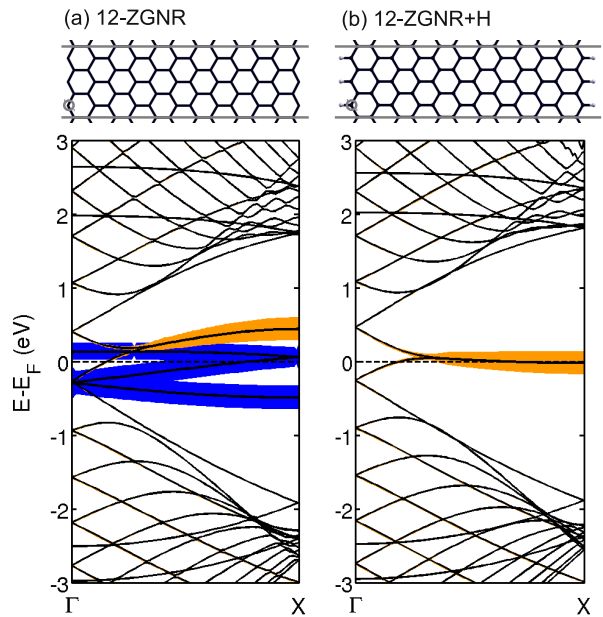}
\caption{(Color online) Atomic structures (top) and nonmagnetic band structures (bottom) of a 12-ZGNR with (a) unpassivated edges and (b) monohydrogenated edges. The fatness of the bands is proportional to the orbital character of the corresponding wave function on the encircled edge atom (top). Orange (light gray) represents the $\pi$ character of the bands, and blue (dark gray) represents their $\sigma$ character. Gray horizontal lines were added to the atomic structure (top), in order to indicate the boundaries of the corresponding unit cell along the periodic direction (see also Fig.~\ref{fig:spindens}).
Because of the presence of (partially) flat bands near $E_\mathrm{F}$ in the nonmagnetic band structure, there will be a magnetic transition into an antiferromagnetic ground state with a band gap (not shown).
}
\label{fig:bands_12-ZGNR}
\end{figure}

\begin{figure}[tb]
\includegraphics[width=\columnwidth,angle=0]{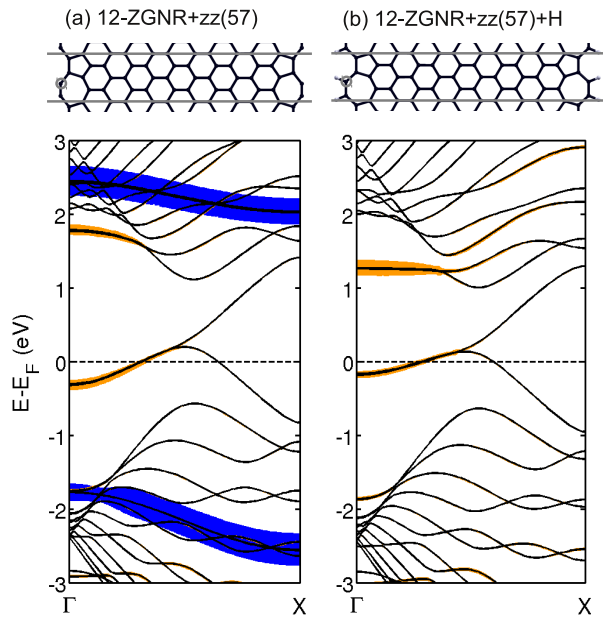}
\caption{(Color online) Atomic structures and band structures of a 12-ZGNR with a $zz(57)$ edge reconstruction for (a) unpassivated edges, (b) monohydrogenated edges (see Fig.~\ref{fig:bands_12-ZGNR}). The system is metallic, nonmagnetic, and has edge states. }
\label{fig:bands_zz(57)}
\end{figure}

\begin{figure}[tb]
\includegraphics[width=.6\columnwidth,angle=0]{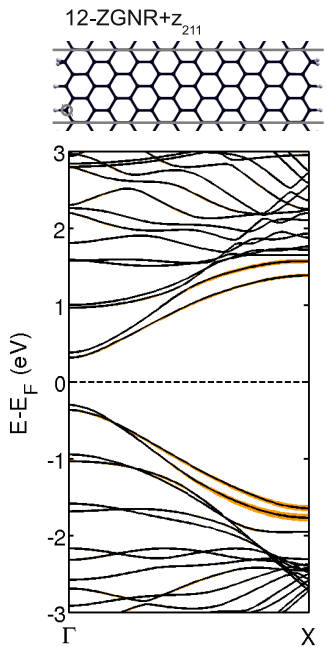}
\caption{(Color online) Atomic structure and band structure of a 12-ZGNR with a $z_{211}$ edge (see Fig.~\ref{fig:bands_12-ZGNR}). The system does not have edge states, it is nonmagnetic, and a semiconductor with a band gap of 0.61 eV.}
\label{fig:bands_z211}
\end{figure}

In our study we consider 10-ZGNRs and 12-ZGNRs, where the prefix `10' or `12' represents the number of zigzag lines across the width of the ribbon. A suffix `+H' indicates that each carbon edge atom is passivated by one hydrogen atom, `+EV' that one edge contains an edge vacancy, `+$zz(57)$' that the edge is reconstructed with alternating lines of pentagons and heptagons, and `+$z_{211}$' that one out of three edge atoms is passivated by two hydrogen atoms and the other two by one hydrogen atom.
The atomic structures of some of these systems are shown in Figs.~\ref{fig:bands_12-ZGNR}-\ref{fig:spindens}.
For the calculations of a H$_2$ molecule, graphene and the ZGNRs, we simulate free-standing objects, where the vacuum region between replica in neighboring unit cells is at least 10 {\AA} wide.
\footnote
{
The optimized lattice parameters of the corresponding unit cells in their (magnetic) ground states are given as follows:
graphene: A=B=2.481, C=10 \AA,
H$_2$: A=B=C=10 \AA,
bcc iron: A=2.793 {\AA} (cubic lattice constant), 
NiO: A=2.932, C=14.326 {\AA} (magnetic unit cell in the hexagonal setting).
For all ZGNRs B=38$\dots$39 \AA, C=10 \AA,
10-ZGNR: A=7.416 \AA,
10-ZGNR+H: A=7.383 \AA,
12-ZGNR: A=7.411 \AA,
12-ZGNR+H: A=7.383 \AA,
10-ZGNR+EV: A=7.399 \AA,
10-ZGNR+EV+H: A=7.344 \AA,
12-ZGNR+$zz(57)$: A=4.899 \AA,
12-ZGNR+$zz(57)$+H: A=4.874 \AA,
12-ZGNR+$z_{211}$: A=7.411 \AA.
}

In Table \ref{tab:edge_energies} we list the edge energies of all ZGNRs that are considered in this study. The edge energy quantifies the energy needed to form an edge from infinite graphene and (for hydrogen passivated systems) molecular hydrogen, i.e. the enthalpy of the virtual reaction
\begin{equation*}
\mbox{graphene} + N_\mathrm{H}/2 \ \mbox{H}_2 \longrightarrow \mbox{ZGNR}.
\end{equation*}
The edge energy can be defined per edge atom ($E_\mathrm{edge}^\mathrm{at}$) or per unit length along the edge ($E_\mathrm{edge}^\mathrm{len}$) as follows
\begin{eqnarray}
E_\mathrm{edge}^\mathrm{at} &=& ( E_\mathrm{tot}^\mathrm{ZGNR} - N_\mathrm{C} E_\mathrm{coh}^\mathrm{graphene} - N_\mathrm{H} E_\mathrm{coh}^\mathrm{H_2} )/N_\mathrm{C}^\mathrm{edge}
\label{eqn:E_edge_at}\\
E_\mathrm{edge}^\mathrm{len} &=& ( E_\mathrm{tot}^\mathrm{ZGNR} - N_\mathrm{C} E_\mathrm{coh}^\mathrm{graphene} - N_\mathrm{H} E_\mathrm{coh}^\mathrm{H_2} )/2A
\label{eqn:E_edge_len}
\end{eqnarray}
where $N_\mathrm{C}$, $N_\mathrm{H}$, and $N_\mathrm{C}^\mathrm{edge}$ are the number of carbon, hydrogen, and carbon edge atoms per unit cell, respectively. $E_\mathrm{tot}^\mathrm{ZGNR}$  is the total energy per unit cell of the ZGNRs in the (magnetic) ground state as given in Tab.~\ref{tab:mag_energies}, $E_\mathrm{coh}^\mathrm{graphene}$ = -9.244 eV/atom and $E_\mathrm{coh}^\mathrm{H_2}$ = -3.394 eV/atom are the cohesive energies of a carbon atom in graphene and a hydrogen atom in the H$_2$ molecule, respectively. $A$ is the lattice constant in the periodic direction, as indicated in Fig.~\ref{fig:spindens}(b). Note that in Eq.~\ref{eqn:E_edge_len} we divided by $2A$, because the unit cell contains both edges of the ZGNR. Our results for edge energies compare well with previous data for such systems. \cite{koskinen_2008_prl,wassmann_2008_prl}

The hydrogen adsorption energies in Tab.~\ref{tab:edge_energies} quantify the gain in energy of a hydrogen passivated ZGNR, as compared to an unpassivated system and molecular hydrogen. In other words, it is the enthalpy of the reaction
\begin{equation*}
\mbox{ZGNR} + N_\mathrm{H}/2 \ \mbox{H}_2 \longrightarrow \mbox{`ZGNR+H'}.
\end{equation*}
The hydrogen adsorption energy per carbon edge atom is defined as
\begin{eqnarray}
E_\mathrm{ads}^\mathrm{H} = (E_\mathrm{tot}^\mathrm{ZGNR+H} - E_\mathrm{tot}^\mathrm{ZGNR} - N_\mathrm{H} E_\mathrm{coh}^\mathrm{H_2})/N_\mathrm{C}^\mathrm{edge},
\label{eqn:E_ads}
\end{eqnarray}
where $E_\mathrm{tot}^\mathrm{ZGNR+H}$ and $E_\mathrm{tot}^\mathrm{ZGNR}$ are the total energies of the hydrogen passivated and the non-passivated ZGNR, respectively.

For magnetic systems we performed collinear, spin-polarized DFT calculations. The results are listed in Tab.~\ref{tab:mag_energies}, comprising energies and magnetic moments for nonmagnetic (NM), ferromagnetic (FM), and antiferromagnetic (AFM) states.
The energies of different magnetic states are compared via the quantity $\Delta E_\mathrm{mag}$, that is the difference in total energy with respect to the magnetic ground state ($E_\mathrm{tot}^\mathrm{GS}$) per magnetic atom:
\begin{eqnarray}
\Delta E_\mathrm{mag} = (E_\mathrm{tot} - E_\mathrm{tot}^\mathrm{GS})/N_\mathrm{MA},
\label{eqn:E_mag}
\end{eqnarray}
where $N_\mathrm{MA}$ is the number of magnetic atoms per unit cell. For ZGNRs $N_\mathrm{MA}$ is the number of magnetic carbon edge atoms per unit cell.

\begin{table}[t]
\begin{tabular}{|l|c|c|c|c|c|c|c|c|c|c|}
\hline
System  & $E_\mathrm{edge}^\mathrm{at}$ & $E_\mathrm{edge}^\mathrm{len}$ & $E_\mathrm{ads}^\mathrm{H}$\\
& (eV/at) & (eV/{\AA}) & (eV/at)\\
\hline \hline
10-ZGNR 	   & 2.99 & 1.21 & --\\
10-ZGNR+H          & 0.26 & 0.11 & -2.73\\
12-ZGNR 	   & 3.00 & 1.21 & --\\
12-ZGNR+H          & 0.28 & 0.11 & -2.72\\
12-ZGNR+$z_{211}$  & 0.08 & 0.03 & -2.91\\
12-ZGNR+$zz(57)$   & 2.39 & 0.97 & --\\
12-ZGNR+$zz(57)$+H & 0.87 & 0.36 & -1.52\\
10-ZGNR+EV         & 2.88 & 1.17 & --\\
10-ZGNR+EV+H       & 0.86 & 0.35 & -2.02\\
\hline
\end{tabular}
\caption{\label{tab:edge_energies}
Edge energies $E_\mathrm{edge}^\mathrm{at}$ and $E_\mathrm{edge}^\mathrm{len}$ as defined by equations \ref{eqn:E_edge_at} and \ref{eqn:E_edge_len},  as well as hydrogen adsorption energies $E_\mathrm{ads}^\mathrm{H}$ as defined by equation \ref{eqn:E_ads}, both calculated at the DFT/GGA/PAW level of theory.
}
\end{table}

\section{Discussion}
\subsection{Stability of edge states}

In a planar $sp^2$ system like graphene, the electronic states split into in-plane ($\sigma$) and out-of-plane ($\pi$) states that are decoupled by symmetry. At an unpassivated zigzag edge the hexagonal carbon network is interrupted, and both the $\sigma$ and the $\pi$ system form edge states. The edge states of the $\sigma$ system are unpaired electrons in $sp^2$ orbitals, i.e. dangling $\sigma$ bonds. The edge states of the $\pi$ system are the ones that were introduced above, and those are usually discussed in the literature. Within the nonmagnetic, electronic band structure of an unpassivated ZGNR both sets of edge states appear as nearly flat bands (or as flat parts of bands) at $E_\mathrm{F}$. \cite{kawai_2000_prb,koskinen_2008_prl} These bands are highlighted in Figs.~\ref{fig:bands_12-ZGNR}-\ref{fig:bands_z211}. In all of these plots, the fatness of the bands is proportional to the orbital character of the corresponding wave function at an edge atom. Orange (light gray) represents their $\pi$ character, and blue (dark gray) represents their $\sigma$ character. The nonmagnetic band structure of an unpassivated 12-ZGNR is shown in Fig.~\ref{fig:bands_12-ZGNR}(a). Here the $\sigma$ and $\pi$ edge states exist in a range of about 0.5 eV above and below $E_\mathrm{F}$. Only the degenerate part of the $\pi$ band has edge state character, and its non-degenerate part has bulk state character.
Flat bands lead to a high DOS. As the chemical reactivity is proportional to the DOS near $E_\mathrm{F}$, an unpassivated ZGNR would be highly reactive, since the edge states are available to form chemical bonds. In other words, the edge states do not only lead to a magnetic instability, 
but they also lead to a chemical and structural instability. The latter will be discussed below. For the 12-ZGNR the magnetic instability leads to an AFM ground state with a band gap of 0.7 eV (Fig.~\ref{fig:bands_12-ZGNR}(a) shows the band structure of the metastable NM state).
The high instability of an unpassivated edge is marked by a rather high edge energy $E_\mathrm{edge}^\mathrm{len}$ of the 10-ZGNR and 12-ZGNR systems of 1.21 eV/{\AA} (see Tab.~\ref{tab:edge_energies}).

\subsubsection{Edge passivation}
One stabilization mechanism for an unpassivated ZGNR is edge passivation. The system will catch any bond partner it can possibly get to saturate its dangling bonds. The majority of papers on ZGNRs consider mono-hydrogenated edges with a single hydrogen atom per edge atom. In these systems the dangling $\sigma$ bonds are saturated, and the $\sigma$ edge states are removed from the vicinity of $E_\mathrm{F}$ [see Fig.~\ref{fig:bands_12-ZGNR}(b)]. However, the $\pi$ edge states persist and render the system magnetic, which results in an AFM ground state discussed above (the band structure of the AFM  state is not shown). For the 10-ZGNR and 12-ZGNR the edge energy drops from 1.21 eV/{\AA} to 0.11 eV/{\AA} after passivation, which stabilizes these systems rather significantly.

However, this type of edge passivation does not correspond to the thermodynamic ground state, as shown by Wassmann \textit{et al.} on the basis of  DFT/GGA calculations. \cite{wassmann_2008_prl} According to these authors the mono-hydrogenated zigzag edge (labeled $z_1$ in their paper), as well as the $zz(57)$ edges discussed below, are only stable at extremely low hydrogen concentrations ($P  \leqslant 10^{-20}$ bar at room temperature). At realistic hydrogen pressures, the only stable zigzag edge should be the $z_{211}$ state shown in Fig.~\ref{fig:bands_z211}, with two mono- and one dihydrogenated sites.
Our calculation on a 12-ZGNR with the $z_{211}$ edge confirm these findings, as the corresponding edge energy is only 0.03 eV/{\AA}, i.e.\ more than a factor of three lower than the monohydrogenated edges.
The resulting system does not have any edge states. It is a semiconductor with a band gap of 0.61 eV, and it is nonmagnetic (see Fig.~\ref{fig:bands_z211}). The $z_{211}$ edge is by far the most stable zigzag edge that is known up to now.

For the $z_{211}$ passivation type the dihydrogenated edge atoms are $sp^3$ hybridized. Therefore a departure from the regular $sp^2$ structure at the edge will be a perfect way to get rid of the edge states, thus removing the peak within the DOS at $E_\mathrm{F}$ and stabilizing the overall structure.

\subsubsection{Edge reconstruction}
Another stabilization mechanism for an unpassivated ZGNR is a structural deformation via edge reconstruction. In their DFT/GGA calculations Koskinen \textit{et al.}~show that a planar reconstruction with alternating pentagon-heptagon carbon rings along the edge, named $zz(57)$ (see Fig.~\ref{fig:bands_zz(57)}), is 0.35 eV/{\AA} lower in edge energy than the normal zigzag edge. \cite{koskinen_2008_prl}
In our own calculations of a 12-ZGNR+$zz(57)$ system we found that the reconstructed edge is 0.24 eV/{\AA} lower in edge energy than the 12-ZGNR system (see Tab.~\ref{tab:edge_energies}). Compared to the edge energies of the hydrogen passivated systems this is only a modest lowering, indicating that the $zz(57)$ edge reconstruction is likely to occur only in high vacuum.

The Koskinen reconstruction leads to the formation of triple bonds in the nearly linear parts of the heptagons, which shift the $\sigma$ edge states away from $E_\mathrm{F}$ [see Fig.~\ref{fig:bands_zz(57)}(a)].
The $\pi$ edge states remain near $E_\mathrm{F}$ in two normally dispersed bands, that render the system metallic. However, there is no magnetic transition, because these bands are not flat. We confirmed this fact on the basis of spin-polarized calculations, which all converged into a NM state. Again, only the degenerate part of the band at $E_\mathrm{F}$ can be associated with edge states, the non-degenerate part has bulk character.
In the hydrogen passivated 12-ZGNR+$zz(57)$+H the $\sigma$ edge states are removed from the band structure, and the occupied $\pi$ bands are practically unchanged [see Fig.~\ref{fig:bands_zz(57)}(b)].

Because of the absence of $\sigma$ states near $E_\mathrm{F}$ Koskinen \textit{et al.} argue that $zz(57)$ reconstructed edges do not need to be stabilized by passivation. They further show that such systems do not favor hydrogen passivation, and therefore they call the $zz(57)$ reconstruction ``self-passivating''.
Our results support this judgment only partially. First, the edge energy of 0.97 eV/{\AA} for the unpassivated 12-ZGNR+$zz(57)$ is still very high.  And second, we find that the hydrogen adsorption energy is $-1.52$ eV per edge atom. Here a negative sign indicates that hydrogenation leads to an increase of binding energy, i.e. that hydrogen adsorption is favored.
After passivation the edge energy of the system (12-ZGNR+$zz(57)$+H) drops to a value of 0.36 eV/{\AA}, indicating that the passivated system is energetically much more favorable than the unpassivated system.
However, because of the absence of $\sigma$ states near $E_\mathrm{F}$, hydrogenation is the least favorable process for these particular systems, in contrast to all other systems. This is reflected in a lower hydrogenation energy, and a relatively high edge energy of the 12-ZGNR+$zz(57)$+H system.

Among other unusual edge reconstructions, the existence of the $zz(57)$ type was confirmed by analyzing aberration-corrected transmission electron microscopy (TEM) graphs of free-standing graphene. \cite{girit_2009_sci,koskinen_2009_prb}$^,$
\footnote
{
The vast majority of graphs in Ref.~[\onlinecite{girit_2009_sci}] show zigzag edges. But one has keep in mind, that the system in the TEM is not in equilibrium, because the electron beam constantly sputters edge atoms away.
Under such conditions, the zigzag edge is simply the edge that can be repaired most rapidly, f.e.\ by diffusing single carbon atoms. Therefore it appears to be more abundant. However the equilibrium shape of different edges cannot be inferred from this experimental study.
}

\subsubsection{Edge closure}
Finally, edge closure in multi-layered graphene is the third stabilization mechanism we would like to emphasize. Liu \textit{et al.}~study the edge structures of graphite by atomically resolved high-resolution TEM. \cite{liu_2009_prl} In a series of tilting experiments the authors demonstrate, that after thermal treatment (at 2000$^\circ$C)  the zigzag and armchair edges between adjacent graphene layers are mostly closed, and that they tend to form folded layers. Edge closure in multi-layered graphene therefore seems to be a simple way to get rid of edges, as well as of the corresponding edges states [see Fig.~\ref{fig:outline}(c)].

\begin{table}[t]
\begin{tabular}{|l|c|c|c|c|c|c|c|c|c|c|}
\hline
System  & State & $N_\mathrm{at}$ & $\mu$ & $E_\mathrm{tot}$ & $\Delta E_\mathrm{mag}$ \\
&&& ($\mu_\mathrm{B}$) & (eV) & (meV/at) \\
\hline \hline
10-ZGNR & NM  & 60 & 0.0 & -535.04379  & 269 \\
     	& FM  & 60 & 7.7 & -536.62974  & 4 \\
     	& AFM & 60 & 0.0 & -536.64359  & 0 \\
\hline
12-ZGNR & NM  & 72 & 0.0 & -645.97119  & 265  \\ 
     	& FM  & 72 & 7.7 & -647.54796  &  2 \\ 
     	& AFM & 72 & 0.0 & -647.56282  &  0 \\ 
\hline
10-ZGNR+H & NM  & 66 & 0.0 & -573.24160  &  27 \\
     	  & FM  & 66 & 1.4 & -573.36746  &  6  \\
     	  & AFM & 66 & 0.0 & -573.40631  &  0  \\
\hline
12-ZGNR+H
	& NM  & 78 & 0.0 & -684.05304  & 29 \\ 
     	& FM  & 78 & 1.4 & -684.20262  & 4 \\ 
     	& AFM & 78 & 0.0 & -684.22471  & 0 \\ 
\hline
10-ZGNR+EV
	& NM  & 59 & 0.0 & -527.32015  & 260 \\ 
     	& FM  & 59 & 4.0 & -528.09920  & 0\footnote[1]
{Only the defect-free edge carries magnetic moments.}
\\ 
\hline
10-ZGNR+EV+H
	& NM  & 64 & 0.0 & -557.08600  & 33\\ 
     	& FM  & 64 & 1.0 & -557.18430  & 0$^\mathrm{a}$ \\ 
\hline \hline
Fe 	& NM & 1 & 0.0 & -7.7566476 & 395\\
	& FM & 1 & 2.1 & -8.1517557 & 0\\
\hline
NiO	& NM & 12 & 0.0 & -68.678312 & 244\\
	& FM & 12 & 4.8 & -68.722936 & 237\\
	& AFM& 12 & 0.0 & -70.142254 &   0\\
\hline
\end{tabular}
\caption{\label{tab:mag_energies}
Magnetic states of different systems calculated at the DFT/GGA/PAW level of theory.
``State'' denotes states with different magnetic ordering: NM (non-magnetic), FM (ferromagnetic), AFM (antiferromagnetic).
$N_\mathrm{at}$ is the number of atoms per unit cell.
$\mu$ is the total magnetic moment per unit cell.
$E_\mathrm{tot}$ is the total energy per unit cell.
$\Delta E_\mathrm{mag}$ is defined in Eq.~\ref{eqn:E_mag}.
}
\end{table}

\subsection{Stability of edge magnetism}
To summarize the first part of our discussion, we showed that ZGNRs with magnetic edge states are not very likely to exist. Nevertheless let us now hypothesize that such systems could be made, and let us discuss the resulting edge magnetism in more detail throughout the second part of this paper.

We start our discussion with two standard magnetic systems, the FM iron (bcc phase, space group Fm\={3}m) and the AFM NiO (rhombohedral lattice system, space group R\={3}m). The energies and magnetic moments of different magnetic states for the two systems are given in Tab.~\ref{tab:mag_energies}. For both systems the ground state magnetic configuration is separated from other possible magnetic configurations by an energy gap of several hundred meV per magnetic atom. This large energy gap is the reason, why magnetic order persists at room temperature.
The size of this energy gap $\Delta E_\mathrm{mag}^\mathrm{GS+1}$, as given by the value of $\Delta E_\mathrm{mag}$ for the second lowest (GS+1) energy state, is an indication of the thermal stability of the magnetic state.
$\Delta E_\mathrm{mag}^\mathrm{GS+1}$ defines an an upper bound for the critical temperature $T_\mathrm{c}^\mathrm{max}$ for spontaneous magnetization ($\Delta E_\mathrm{mag}^\mathrm{GS+1} = \mathrm{k} T_\mathrm{c}^\mathrm{max}$).
The real critical temperatures are usually much smaller. For iron and NiO the we find $T_\mathrm{c}^\mathrm{max}$ to be about five times bigger than the experimentally measured $T_\mathrm{c}$. Below we will use $T_\mathrm{c}^\mathrm{max}$ to estimate the thermal stability of spontaneous magnetization in ZGNRs.

\subsubsection{Ideal ZGNRs}
\begin{figure}[tb]
\includegraphics[width=\columnwidth,angle=0]{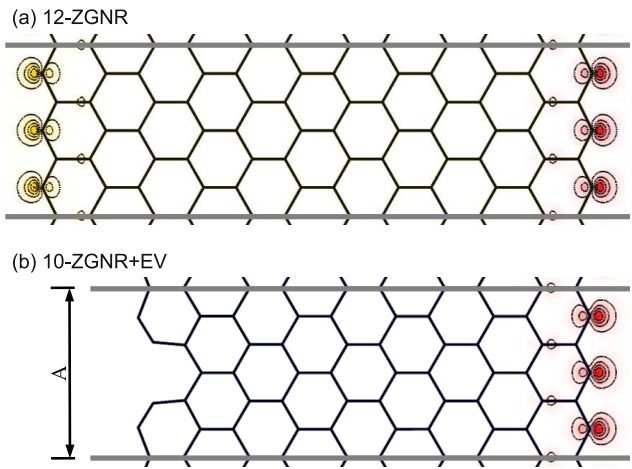}
\caption{(Color online) The spin density of (a) an unpassivated 12-ZGNR in the AFM state and (b) an unpassivated 10-ZGNR with an edge vacancy in the FM state.
The red (dark gray) contours on the right hand side represent positive spin densities within the graphene layer; the yellow (light gray) contours on the left hand side represent negative densities.
$A$ is the lattice constant along the periodic direction. (b) Due to the presence of the defect, there are no remaining magnetic moments on the left edge.}
\label{fig:spindens}
\end{figure}

Now consider the magnetic states of the ideal, unpassivated and monohydrogenated systems 10-ZGNR, 12-ZGNR 10-ZGNR+H, 12-ZGNR+H. The 12 carbon chain systems are shown in Fig.~\ref{fig:bands_12-ZGNR}. Table \ref{tab:mag_energies} indicates that for each of these systems a NM, FM, and AFM state can be found. For the FM configuration all magnetic atoms are aligned along the magnetic axis, and in the AFM configuration the atoms are ferromagnetically ordered along one edge, and antiferromagnetically ordered between opposite edges. Figure \ref{fig:spindens}(a) depicts the spin density of the AFM state of the 12-ZGNR. All of these results are in excellent agreement with other DFT studies of similar systems. \cite{lee_2005_prb,son_2006_prl,martins_2007_prl}

The AFM configuration is the lowest energy state of all systems.
However, the FM states are only $\Delta E_\mathrm{mag}^\mathrm{GS+1}=2-6$ meV per edge atom higher in energy. The upper bound for the critical temperatures in these cases is $T_\mathrm{c}^\mathrm{max}= 70$ K.
Furthermore, these very small energy differences are of the order of the precision of modern DFT calculations. So within DFT the AFM and FM states can hardly be distinguished at all from an energetic point of view. The reason for this finding is the large width of the ribbons (19.8 and 24.1 {\AA} for 10-ZGNR and 12-ZGNR, respectively), which leads to a decoupling of the magnetic edges. This is in perfect agreement with the well known finding, that for large ribbon widths, the FM and AFM states are degenerate. \cite{lee_2005_prb}
For ZGNRs with smaller widths the interactions between the tails of the edge states are more pronounced, and the energy differences between the AFM and FM configurations $\Delta E_\mathrm{mag}^\mathrm{GS+1}$ increase. It has been shown by Jung \textit{et al.} that $\Delta E_\mathrm{mag}^\mathrm{GS+1}$  follows a $C/W^2$ law, where $W$ is the ribbon width and $C=2.7$ eV{\AA}$^2$. \cite{jung_2009_prl} But even for very narrow ZGNRs, $\Delta E_\mathrm{mag}^\mathrm{GS+1}$ will not significantly exceed 30 meV.

We conclude that the edge magnetism in ideal, unpassivated and monohydrogenated ZGNRs will not be stable at room temperature. Therefore practical applications in spintronic devices will not be feasible. We estimate that the AFM state can only be stabilized at temperatures $T \lesssim 10$ K.
Above that temperature ideal ZGNRs will simply exhibit paramagnetic behavior.

These findings are in good agreement with a recent experimental study by Sepioni \textit{et al.} \cite{sepioni_2010_prl} Those authors studied graphene laminates by SQUID magnetometry, and found that the laminates are strongly diamagnetic and exhibit no sign of ferromagnetism down to 2 K. Below 50 K they detect a very weak paramagnetic contribution.
These results can be explained by our previous argument, that the majority of edges is either reconstructed, passivated, or closed. Only a few remaining magnetic edges would give rise to the very weak paramagnetic contribution. This explanation, however, cannot account for very narrow  distribution of magnetic moments that was measured in Ref.~[\onlinecite{sepioni_2010_prl}].

A further confirmation of the highly unstable magnetism in ZGNRs was given by Yazyev \textit{et al.} \cite{yazyev_2008_prl} They used a DFT parametrized spin-Heisenberg model to show, that the spin correlation length of the magnetic edges is only about 1 nm at room temperature. 

\subsubsection{Edge Defects}
Under current experimental conditions one cannot produce such ideal ZGNR. It is also rather questionable, whether this will be possible at all. In reality one always has to deal with non-ideal structures, containing impurities, vacancies, lattice defects, etc.\cite{faccio_2010_jpcc}

Figure \ref{fig:spindens}(b) shows a 10-ZGNR with an edge vacancy (EV) on the left hand side of the ribbon. This vacancy totally suppresses any magnetic moments on that side, but it does not influence the moments on the other side. Therefore only one magnetic solution (FM) exists for 10-ZGNR+EV and 10-ZGNR+EV+H. Apart from that, the values of $\Delta E_\mathrm{mag}$ for the FM an NM states are similar to previous cases 
without edge vacancy. This result shows again that opposite edges are almost decoupled.

For the unpassivated 10-ZGNR+EV $\Delta E_\mathrm{mag}^\mathrm{GS+1}$ is very high.
This is an effect of the dangling $\sigma$ bonds, which have unpaired electrons with individual magnetic moments of about 1 $\mu_\mathrm{B}$. These unpaired electrons render the system strongly magnetic. However, as discussed above, unpassivated systems are highly unstable and they will never exist at ambient conditions. The instability of this system is again reflected by its high edge energy of 1.17 eV/{\AA} (see Tab.~\ref{tab:edge_energies}). It is interesting to notice that the edge energy of the 10-ZGNR+EV is lower than the edge energy of an ideal 10-ZGNR by 0.04 eV/{\AA}. Thus for unpassivated systems the formation of EV is energetically favorable. The reason is that the EV removes $\sigma$ edge states from the vicinity of $E_F$. This finding might be important for the understanding of disorder during the growth process, where unpassivated edges might exist at an early stage.

However, if the edges are passivated, as for the 10-ZGNR+EV+H system, no $\sigma$ edge states are present. Consequently the formation of EV will not be favored any more ($E_\mathrm{edge}^\mathrm{len}$ is 0.21 eV/{\AA} higher than for 10-ZGNR+H). 
As the 10-ZGNR+EV+H system has magnetic moments only along one edge $\Delta E_\mathrm{mag}^\mathrm{GS+1} = 33$ meV can be considered as the energy of moment formation along a single edge. 
The corresponding upper bound for the critical temperature $T_\mathrm{c}^\mathrm{max}= 380$ K is significantly higher than for ideal ZGNRs ($T_\mathrm{c}^\mathrm{max} = 70$ K). However we want to remind the reader that real critical temperatures are usually much lower than $T_\mathrm{c}^\mathrm{max}$. So room temperature applications of edge magnetism along a single edge are also unlikely.

Huang \textit{et al.} systematically studied the effect of edge defects (vacancies and substitutional dopants) on the magnetism with DFT/LDA calculations. \cite{huang_2008_prb} The stability of the magnetism is found to continuously decrease with increasing concentration of the defects. ZGNRs become nonmagnetic at concentrations of about one edge defect impurity per 10 {\AA} (in our example the concentration is one defect per 7.4 {\AA}). The authors judge that such critical defect concentrations are accessible in real samples. The corresponding suppression of edge magnetism is mainly caused by the removal of edge states from $E_\mathrm{F}$.

Yazyev \textit{et al.} used DFT/GGA calculations to show that the magnetic domain wall creation energy of 114 meV for the ideal zigzag edge decreases dramatically in the presence of edges defects, down to energies of the order of 30 meV. This means that magnetic alignment along one edge, which might be relatively stable in ideal ZGNRs, will easily be destroyed by the occurrence of edge defects. \cite{yazyev_2008_prl}

We therefore conclude that the edge magnetism, which is already a very weak effect in ideal ZGNRs, is further weakened or even totally disappears in the presence of edge defects. The stability of the edge magnetism could only be enhanced, if high density edge defects are confined to one edge of the ribbon, while the other edge remains in an ideal zigzag shape (10-ZGNR+EV+H). However, even in this very artificial case, the magnetism is still too weak to allow for room temperature applications.

\subsubsection{Charge doping}
In experimental measurements of graphene systems one always has to deal with (intentional or unintentional) charge doping effects, which are induced by the substrate/back-gate or by various impurities.

\begin{figure}[tb]
\includegraphics[width=0.7\columnwidth,angle=0]{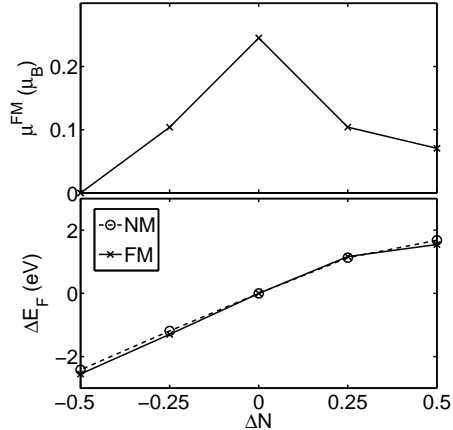}
\caption{The effect of charge doping in 12-ZGNR+H: $\Delta N$ is the number of excess electrons per edge atom, $\mu^\mathrm{FM}$ is the magnetic moment per edge atom in the FM state, and $\Delta E_\mathrm{F}$ is the change of the Fermi energy with respect to the undoped case for both the NM and the FM state.}
\label{fig:doping}
\end{figure}

In Fig.~\ref{fig:doping} we show the effect of charge doping on a 12-ZGNR-H system. $\Delta N$ is the number of excess electrons per edge atom. Positive values represent electron doping, and negative values represent hole doping. We normalize $\Delta N$ to the number of edge atoms, because 
the edge states are the states, which are directly affected by the doping.
We see that progressive charge doping by electrons or holes move $E_\mathrm{F}$ away from the aforementioned peak in the DOS (at $\Delta E_\mathrm{F} = 0$), which gradually suppresses the magnetic transition. Therefore the magnetic moment per edge atom ($\mu^\mathrm{FM}$) becomes smaller and smaller until it vanishes eventually, and the corresponding system becomes NM.
For small values of $\pm \Delta N$ electron and hole doping is symmetric, but for larger values the two scenarios become more and more asymmetric. This is because the band structure of the undoped 12-ZGNR+H system [see Fig.~\ref{fig:bands_12-ZGNR}(b)] is symmetric near $E_\mathrm{F}$, and it is becoming more and more asymmetric away from $E_\mathrm{F}$. Furthermore for large doping levels the rigid band model breaks down, and the band structure of the doped system changes significantly, as compared to the undoped case.

The critical doping level for a 12-ZGNR-H is ca.~0.5 electrons per edge atom, which agrees very well with previous results. \cite{wimmer_2008_prl,jung_2009_prb} However this is a very high doping level, which is not reached in standard experimental settings. The charge fluctuations in typical graphene charge puddles are of the order of $n_\mathrm{2D} = 10^{11}$ cm$^{-2}$, and back-gate doping can inject charge carriers in concentrations up to $n_\mathrm{2D} = 10^{12}-10^{13}$ cm$^{-2}$.  \cite{martin_2008_natphys,zhang_2009_natphys} Such substrate induced charge concentrations ($n_\mathrm{2D} = 10^{11}-10^{13}$ cm$^{-2}$) would correspond to doping levels $\Delta N$ between 0.0006 and 0.06 electrons/holes per edge atom for the 12-ZGNR-H ribbon.\footnote{
The doping level is $\Delta N = n_\mathrm{2D} W A / N_\mathrm{C}^\mathrm{edge}$.
$A$ is the lattice constant in periodic direction, $W$ is the ribbon width, and $N_\mathrm{C}^\mathrm{edge}$ is the number of carbon edge atoms per unit cell.
}
Under such conditions the weakening of the edge magnetism would only be marginal.

Figure \ref{fig:doping} allows for the translation of the critical doping level into a critical shift in $E_\mathrm{F}$, which is about $\pm$2 eV. So by contacting the 12-ZGNR-H at two points along the periodic direction, and by applying critical voltage of 4 eV (shifting the electronic structure by +2 eV on one contact and by -2 eV on the other contact), one would also destroy the edge magnetism close to the contacts. Smaller voltages may not destroy the edge magnetism, but it will be weakened quite significantly. As the charge mobility in graphene is very high, voltages of the order of a few volts can indeed be applied \cite{murali_2009_apl}, which leads to a significant weakening of the edge magnetism.

We thus conclude that under current experimental conditions, back-gate doping and charge puddles do not significantly influence the edge magnetism in ideal ZGNRs of rather small width. However, when contacting ZGNR electrically, and after applying voltages of a few volts, the edge magnetism will be significantly weakened.

\section{Conclusions}  
In summary, we critically discussed the stability of edge states and edge magnetism in ZGNRs.
In the first part of our paper we pointed out that magnetic edge states are not very likely to exist. We showed that there are at least three very natural mechanisms -- edge reconstruction, edge passivation, and edge closure -- which dramatically reduce the effect of edge states in ZGNRs, or even eliminate them completely.
In the second part of our paper, we showed that if ZGNRs with magnetic edge states could be made, the intrinsic magnetism would not be stable at room temperature. Furthermore, charge doping and the presence of edge defects further destabilize this already rather weak type of collective magnetism.
We conclude that antiferromagnetic ZGNRs might only be observed under ultra-clean, low-temperature conditions in defect-free samples. The present state of research indicates that edge magnetism within graphene ZGNRs is much too weak to be of practical significance, in particular for spintronics applications.

\section{Acknowledgments}
The computations were performed at \c{C}ankaya University and at The Center for Information Services and High Performance Computing (ZIH) of the TU Dresden. The work was supported by DFG SPP 1459. JK and AQ also like to thank G. Cuniberti (TU Dresden) for support.

\bibliographystyle{apsrev4-1}
\bibliography{references}

\end{document}